# Efficient analysis and design of low-loss whispering-gallery-mode coupled resonator optical waveguide bends

Svetlana V. Pishko, Phillip Sewell, *Senior Member, IEEE*, Trevor M. Benson, *Senior Member, IEEE*, and Svetlana V. Boriskina, *Senior Member*, *IEEE*

*Abstract*— Waveguides composed of electromagnetically-coupled optical microcavities (coupled resonator optical waveguides or CROWs) can be used for light guiding, slowing and storage. In this paper, we present a two-dimensional analysis of finite-size straight and curved CROW sections based on a rigorous Muller boundary integral equations method. We study mechanisms of the coupling of whispering gallery (WG) modes and guiding light around bends in CROWs composed of both identical and size-mismatched microdisk resonators. Our accurate analysis reveals differences in WG modes coupling in the vicinity of bends in CROWs composed of optically-large and wavelength-scale microcavities. We propose and discuss possible ways to design low-loss CROW bends and to reduce bend losses. These include selecting specific bend angles depending on the azimuthal order of the WG mode and tuning the radius of the microdisk positioned at the CROW bend.

*Index Terms*— **optical microcavities, microdisk resonators, coupled resonator optical waveguides, photonic molecules, whispering-gallery modes, bend losses, integral equations.**

## I. INTRODUCTION

OPTICAL micro-resonators and coupled-resonator structures open amazing opportunities in many diverse scientific and technological areas such as cavity quantum electrodynamics, non-linear optics, data storage, bio-(chemical)sensing, microlasers, add-drop filters etc [1-7]. Linear chains of side-coupled resonators can also be used for optical power transfer. This novel type of optical waveguide has recently been proposed [8] and then demonstrated and studied in a variety of configurations, such as sequences of microdisks or microrings [9-11], arrays of coupled microspheres [12-15], and chains of Fabry-Perot or photonic crystal defect cavities [9, 16-18]. Among the advantages offered by CROWs are the possibility of making loss-less waveguide bends [8, 15, 17] and significant slowing of light pulses [3, 9, 19, 20].

Manuscript received January, 2007. This work has been partially supported by the NATO Collaborative Linkage Grant CBP.NUKR.CLG 982430. S. V. Pishko and S. V. Boriskina are with the School of Radiophysics, V. Karazin Kharkov National University, Kharkov 61077, Ukraine (fax: 1-831-308-7657; e-mail: SBoriskina@gmail.com). T. M. Benson and P. Sewell are with the George Green Institute for Electromagnetics Research, University of Nottingham, Nottingham NG7 2RD, UK.

Very efficient or even complete transmission through bends in CROWs has been predicted in the pioneering paper [8]. It was suggested that if a resonator mode possesses an *n*-fold rotational symmetry, perfect transmission can be achieved through $2\pi/n$ bends. This statement is partially supported by the results of Ref [15], where efficient transmission through a bent chain of microspheres with a bend angle of 90° has been observed. However, such a prediction is based on the assumption that resonators are weakly-coupled, and thus the mode field patterns in resonators remain essentially the same as in the isolated microcavity. The continued drive for the miniaturization and dense packaging of photonic components has led to the development of ultra-small resonators made of high-index-contrast materials. To achieve efficient evanescent coupling between such resonators they should be brought very close to each other, which may significantly disturb WG-mode field patterns [6, 7, 21]. Furthermore, in the region of a sharp CROW bend strong evanescent-field coupling may occur between non-neighboring resonators. These effects may render the tight-binding approximation used in [8] and many later works inapplicable.

To study in detail the mechanisms of WG modes coupling and forming lossless CROW bends, we performed a comprehensive numerical study of finite-size CROW sections of arbitrary geometry, which is based on a rigorous Muller boundary integral equations (MBIEs) formulation [6, 7, 22]. This formulation reduces the problem space to the resonators surfaces (thus drastically reducing the numerical effort), automatically imposes the radiation condition at infinity, and enables the treatment of both high and low index-contrast materials with material losses and gain. The MBIEs-based algorithm accounts for all the electromagnetic interactions within the system and thus provides superior accuracy of the numerical solutions for both weakly- and strongly-coupled resonators.

Furthermore, the developed algorithm provides us with an efficient and flexible tool for the simulation of aperiodic finite-size CROW sections composed of resonators of different radii. In most studies of CROWs performed to date infinite periodic structures composed of identical resonators have been considered. It should be noted that practical CROWs are always of finite size, and thus their characteristics can differ from those theoretically predicted for infinite structures [23]. Previous research effort directed towards understanding how disorder in either the resonators size or the inter-resonator coupling coefficients affects CROW transmission characteristics has been primarily focused on the estimation of CROW fabrication tolerances [13, 24]. The purpose of this study is completely different; we use the resonators sizes as parameters that can be adjusted to tune the CROW characteristics. In the following sections we show that by properly choosing the size of individual resonators it is possible to achieve very efficient transmission through arbitrary CROW bends.

## II. PROBLEM FORMULATION AND SOLUTION

A finite-size CROW section composed of $L$ side-coupled microdisk resonators is considered as illustrated in Fig. 1. Dielectric (or semiconductor) microdisks of radii $a_l$ and permittivities $\varepsilon_l$ ($l = 1, 2 \dots L$) are located in a host medium with permittivity $\varepsilon_e$. Note that the disks can be of different radii and permittivities and can be arranged into arbitrary 2-D spatial structures.

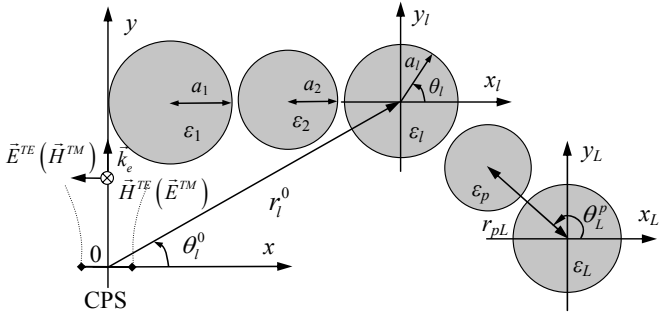

Fig. 1. Curved aperiodic CROW section consisting of microdisk resonators of different radii together with the global and local coordinate systems used in the analysis. Directional beam generated by a line source with a complex coordinate is grazing the rim of the left resonator.

Depending on the polarization, the total field can be uniquely determined from one component, $H_z$ (for TE or transverse-electric modes) or $E_z$ (for TM or transverse-magnetic modes), respectively. Throughout the paper, the time-dependence convention $\exp\{-i\omega t\}$ is adopted and omitted.

The CROW section is excited by the 2-D complex-point source (CPS) beam. CPS is a line source with complex coordinates: $U^{inc} = H_0^{(1)}\left(k\left|\vec{r} - \vec{r}_{cs}\right|\right)$, $\vec{r}_{cs} = \{x_{cs}, y_{cs}\} = \vec{r}_0 + i\vec{b}$, $\vec{r}_0 = \{x_0, y_0\}$, $\vec{b} = \{b\cos\varphi_{cs}, b\sin\varphi_{cs}\}$. Such a source produces a beam field in real space [25]:

$$U^{inc}(\vec{r}) = H_0^{(1)}\left(k_e\left|\vec{r} - \vec{r}_{cs}\right|\right)$$
$$= \begin{cases} \sum_{(n)} J_n(k_e r_{cs}) H_n^{(1)}(k_e r) e^{-in\varphi_{cs}} e^{in\varphi}, & r_{cs} < r \\ \sum_{(n)} J_n(k_e r) H_n^{(1)}(k_e r_{cs}) e^{-in\varphi_{cs}} e^{in\varphi}, & r_{cs} > r \end{cases} \quad (1)$$

Applying the Green's formula to the fields and the Green's functions in the regions inside and outside all the resonators and taking into account the boundary conditions, we can formulate the problem in terms of the 2-D Muller boundary integral equations [6, 22]. Then, using angular exponents $\left\{e^{imt}\right\}_{m=-\infty}^{\infty}$ as global basis and trial functions in the Galerkin scheme, we discretize the MBIEs and obtain the following block-matrix equation of the Fredholm second kind [6]:

$$a_m^p u_m^p + b_m^p v_m^p + \sum_{l \neq p}\left\{ \sum_{(n)} u_n^l A_{mn} + \sum_{(n)} v_n^l B_{mn} \right\} = e_m^p, \quad (2)$$

$$c_m^p u_m^p + d_m^p v_m^p + \sum_{l \neq p}\left\{ \sum_{(n)} u_n^l C_{mn} + \sum_{(n)} v_n^l D_{mn} \right\} = \frac{1}{k} f_m^p, \quad (3)$$

where

$$a_m^p = \sqrt{\varepsilon_p} J_m(k_p a_p) H_m^{(1)'}(k_p a_p)$$
$$- \sqrt{\varepsilon_e} J_m'(k_e a_p) H_m^{(1)}(k_e a_p) + \frac{4}{i\pi k a_p} \quad (4)$$

$$b_m^p = J_m(k_p a_p) H_m^{(1)}(k_p a_p) - \frac{\alpha_e}{\alpha_p} J_m(k_e a_p) H_m^{(1)}(k_e a_p) \quad (5)$$

$$c_m^p = \varepsilon_e J_m'(k_e a_p) H_m^{(1)'}(k_e a_p) - \varepsilon_p J_m'(k_p a_p) H_m^{(1)'}(k_p a_p) \quad (6)$$

$$d_m^p = \frac{\alpha_e}{\alpha_p} \sqrt{\varepsilon_p} J_m(k_e a_p) H_m^{(1)'}(k_e a_p)$$
$$- \sqrt{\varepsilon_p} J_m'(k_p a_p) H_m^{(1)}(k_p a_p) + \frac{2(\alpha_p + \alpha_e)}{i\pi\alpha_p k a_p} \quad (7)$$

and

$$A_{mn} = \left( \sqrt{\varepsilon_l} J_n'(k_l a_l) J_m(k_l a_p) H_{m-n}^{(1)}(k_l r_{pl}) \right.$$
$$\left. - \sqrt{\varepsilon_e} J_n'(k_e a_l) J_m(k_e a_p) H_{m-n}^{(1)}(k_e r_{pl}) \right) e^{i(n-m)\theta_p^l} \quad (8)$$

$$B_{mn} = \left( J_n(k_l a_l) J_m(k_l a_p) H_{m-n}^{(1)}(k_l r_{pl}) \right.$$
$$\left. - \frac{\alpha_e}{\alpha_l} J_n(k_e a_l) J_m(k_e a_p) H_{m-n}^{(1)}(k_e r_{pl}) \right) e^{i(n-m)\theta_p^l} \quad (9)$$

$$C_{mn} = \left( \varepsilon_e J_n'(k_e a_l) J_m'(k_e a_p) H_{m-n}^{(1)}(k_e r_{pl}) \right.$$
$$\left. - \varepsilon_l J_n'(k_l a_l) J_m'(k_l a_p) H_{m-n}^{(1)}(k_l r_{pl}) \right) e^{i(n-m)\theta_p^l} \quad (10)$$

$$D_{mn} = \left( \frac{\alpha_e}{\alpha_l} \sqrt{\varepsilon_e} J_n(k_e a_l) J_m'(k_e a_p) H_{m-n}^{(1)}(k_e r_{pl}) \right.$$
$$\left. - \sqrt{\varepsilon_l} J_n(k_l a_l) J_m'(k_l a_p) H_{m-n}^{(1)}(k_l r_{pl}) \right) e^{i(n-m)\theta_p^l} \quad (11)$$

In (5)-(11), $k$ is the free-space wavenumber, $k_e = k\sqrt{\varepsilon_e \mu_e}$, $k_l = k\sqrt{\varepsilon_l \mu_l}$, $l = 1...L$; $J_m(\cdot)$ and $H_m^{(1)}(\cdot)$ are the Bessel and Hankel functions, respectively, (prime denotes differentiation with respect to the argument); $\alpha_l^{TE} = \varepsilon_l$, $\alpha_e^{TE} = \varepsilon_e$, $\alpha_l^{TM} = \mu_l$, $\alpha_e^{TM} = \mu_e$, $l = 1...L$. Coefficients $a_m - d_m$ correspond to the matrix coefficients of the scattering problem for an isolated $p$-th resonator, while coefficients $A_{mn} - D_{mn}$ describe the optical coupling between the $p$-th and the $l$-th microdisks. It should be noted here that a similar theoretical formalism has been previously applied to study optical properties of photonic crystals consisting of a finite number of infinite cylinders [26-28]. The right-hand side vector of Eqs. (2)-(3) can be calculated as follows:

$$e_m^p = \frac{1}{2\pi} \int_0^{2\pi} U^{inc}(r_p, \theta_p) e^{-im\theta_p} d\theta_p \tag{12}$$
$$= J_m(k_e a_p) e^{-im\theta_p^0} \cdot \sum_{(n)} J_n(k_e r_{cs}) H_{m-n}^{(1)}(k_e r_p^0) e^{-in\varphi_{cs}} e^{in\theta_p^0}$$

$$f_m^p = \frac{\partial e_m^p}{\partial n_p} = -\frac{1}{2\pi} \int_0^{2\pi} \frac{\partial U^{inc}(r_p, \theta_p)}{\partial r_p} e^{-im\theta_p} d\theta_p \tag{13}$$
$$= -k_e J_m'(k_e a_p) e^{-im\theta_p^0} \cdot \sum_{(n)} J_n(k_e r_{cs}) H_{m-n}^{(1)}(k_e r_p^0) e^{-in\varphi_{cs}} e^{in\theta_p^0}$$

The far-field scattering pattern is evaluated by applying the steepest-descent method in the far zone. This brings us to the following expression:

$$\Psi(\varphi) = \sum_{l=1}^{L} \left\{ \sum_{(m)} \sum_{(n)} \left[ \begin{array}{c} \left( \dfrac{\alpha_e}{\alpha_l} v_m^l - \dfrac{i\pi}{2} v_m^0 \right) J_m(k_e a_l) \\ + \sqrt{\varepsilon_e} \left( u_m^l - \dfrac{i\pi k}{2} u_m^0 \right) J_m'(k_e a_l) \end{array} \right] \right. \tag{14}$$
$$\left. \cdot (-i)^n J_{n-m}(k_e r_l^0) e^{i(m-n)\theta_l^0} e^{in\varphi} \right\}$$

The total power scattered from the CROW section can be calculated by evaluating numerically the following integral over a closed contour in the far-zone of the resonators:

$$P_{sc} = \frac{2}{\pi} \int_0^{2\pi} |\Psi(\varphi)|^2 d\varphi \tag{15}$$

We normalize this total scattered power by the value of total power radiated by the CPS in the free space: $P_0 = 4 I_0 (2kb)$, where $I_0(\cdot)$ is the zeroth-order modified Bessel function of the first kind. Note that if there are no additional field sources and all the resonators are made of lossless material, the value of the normalized total scattered power equals one.

In the following sections, we restrict our analysis to the TM-polarization case as mechanisms of TE- and TM-polarized WG-modes coupling and degeneracy splitting in coupled-resonator systems have many similar features.

## III. TRANSMISSION CHARACTERISTICS OF A STRAIGHT FINITE-SIZE CROW SECTION

First, the developed algorithms are applied to study transmission spectra of a straight finite-length CROW section composed of identical WG-mode microdisk resonators. The CROW is excited by the CPS beam grazing the rim of the left resonator in the chain. To detect the energy transport along the whole CROW section, we assume that all the resonators in the CROW apart from the last (right) one have no material losses and look for minima in the normalized total scattered power plots. This procedure is similar to the one used in [29], where the energy transport along metal nanoparticles plasmon waveguides was probed by using fluorescent nanospheres; only we use absorption losses in the last resonator as an evidence of the energy transport along the CROW.

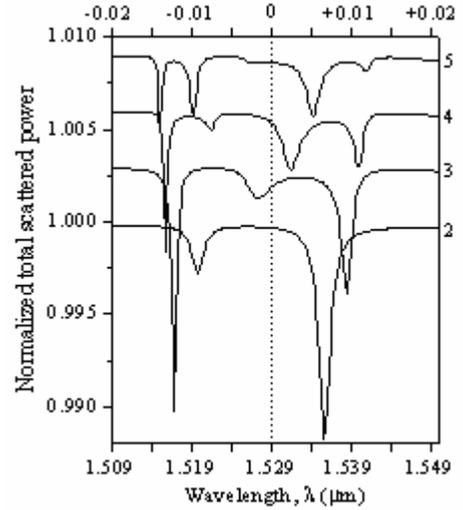

Fig. 2. Normalized total scattered power of a CPS beam scattering from a straight CROW section composed of identical microdisks with $a_i = 0.9 \,\mu\text{m}$, $\varepsilon_i = 7$, $i = 1...L - 1$, $\varepsilon_L = (7, 0.001)$ as a function of the incident field wavelength for varying number of microdisks.

Several frequency scans of the normalized total scattered power of the CPS beam are shown in Fig. 2 for CROW sections composed of two, three, four and five resonators side-coupled via equal airgaps. Here and in the following figures, successive plots are shifted upward for clarity. In Fig. 2, the minima correspond to the frequencies of the high-Q CROW WG$_{7,1}$-supermodes, at which efficient light transmission along the CROW occurs. The dotted line indicates the value of the resonant wavelength of the WG$_{7,1}$-mode in an isolated resonator, and the labels on the top axis indicate the wavelength shift from this value. An increase in the number of coupled cavities leads to the appearance of additional dips, which eventually transform into a broad unstructured band in the spectrum of an infinite CROW if the number of resonators is increased further [16, 30]. Fig. 3 demonstrates the effect of the inter-cavity distance on the CROW transmission spectrum. It can be seen that bringing resonators closer to each other shifts the super-modes wavelength away from the single-cavity WG$_{7,1}$-mode wavelength. Furthermore, narrowing the airgaps reduces the number of dips in the plots.

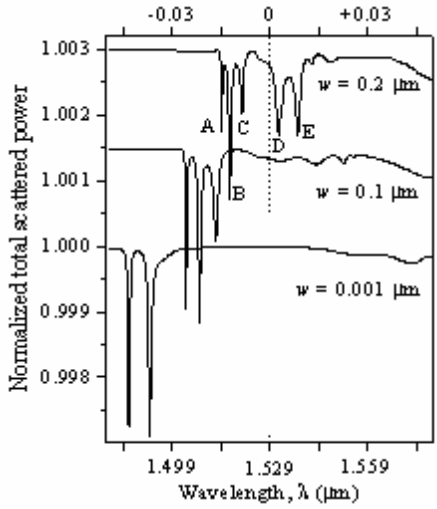

Fig. 3. Normalized total scattered power of a CPS beam scattering from a straight CROW section composed of seven identical microdisks with $a_i = 0.9$ μm, $\varepsilon_i = 7$, $i = 1,...6$, $\varepsilon_7 = (7, 0.001)$ as a function of wavelength for several values of the disk-to-disk airgap width.

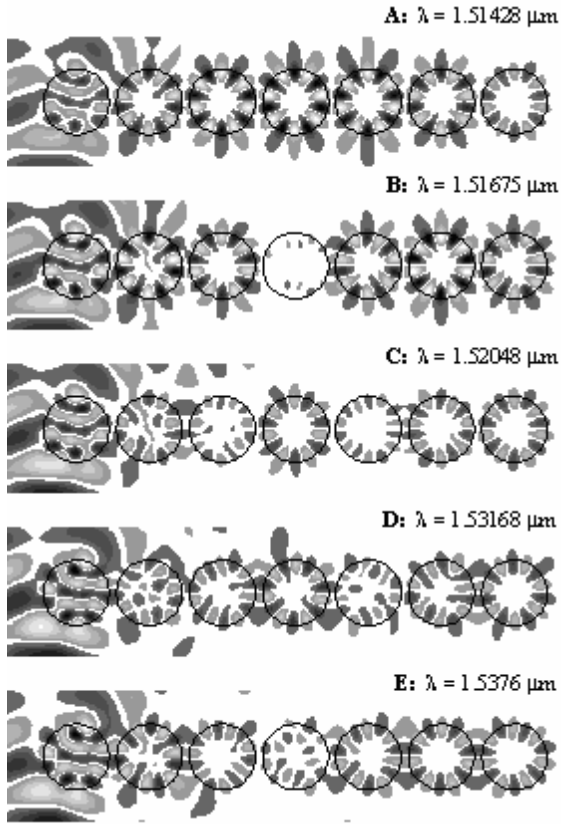

Fig. 4. The electric field profiles of the CROW WG$_{7,1}$-supermodes at the wavelengths corresponding to the total scattered power minima observed in Fig. 3 for $w = 0.2$ μm. The minima are marked from left to right (mode A has the shortest wavelength).

We should now recall the main properties of the WG-supermodes in linear chains of evanescently-coupled microdisk resonators. In the spectrum of a CROW composed of $L$ microdisks, in place of every double-degenerate WG-mode of a single disk $L$ pairs of nearly-degenerate supermodes belonging to various symmetry classes arrear. Bonding WG-supermodes shift to longer wavelengths and anti-bonding supermodes shift to shorter wavelengths [21, 31].

Clearly, the number of deep minima in the transmission characteristics in Fig. 3 is less than the number of supermode doublets expected in the 7-resonator CROW spectrum, especially for narrow airgaps. The reason for this is that efficient light transport along the CROW is only possible on the wavelengths corresponding to the high-Q WG-supermodes. Bringing resonators close to each other may significantly suppress Q-factors of many CROW supermodes [6, 7, 21]. By studying the near-field portraits of the supermodes corresponding to the dips in the CROW transmission characteristics (Fig. 4) we can see that the most efficient light transfer occurs on blue-shifted anti-bonding modes A and B. These modes are known to have the highest Q-factors among all the supermodes of a straight CROW.

## IV. TRANSMISSION THROUGH CROW BENDS

Next, transmission characteristics of curved CROW sections will be studied with the aim to identify general rules to design low-less bends in coupled-cavity waveguides. In Fig. 5, we present the 7-microdisk CROW transmission spectra for several CROW bend angles ranging from 0 to 90 degrees with 10-degree increment and for two values of the disk-to-disk airgap width. The CROW section is bent around the central resonator, and the bend angle is measured from the $x$-axis. By comparing the data presented in Fig.5(a) and Fig.5(b) we can conclude that it may be difficult to achieve efficient light transport through curved CROWs composed of touching wavelength-scale microdisks. Even in the case when the resonators are separated by relatively wide airgaps, most dips that are well-pronounced in the straight CROW spectrum disappear if the CROW bend angle is increased.

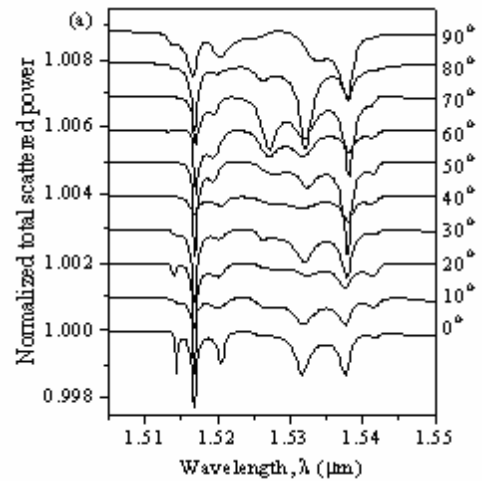

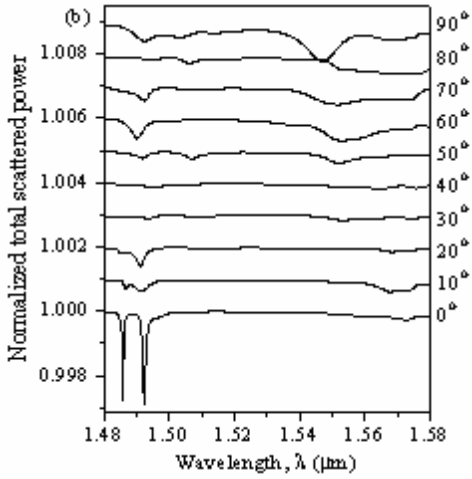

Fig. 5. Normalized total scattered power of a CPS beam scattering from a bent CROW section composed of seven microdisks with $a_i = 0.9\ \mu m$, $\varepsilon_i = 7$, $i = 1,...6$, $\varepsilon_7 = (7, 0.001)$ as a function of wavelength for several values of the CROW bend angle: (a) $w = 0.2\ \mu m$, (b) $w = 0.001\ \mu m$.

This result is in contrast with that predicted for the CROWs composed of weakly-coupled optically large resonators. To understand the mechanism of WG modes coupling in the bent CROW section, we will now study in detail how bending the CROW affects the complex frequency of the CROW natural supermodes corresponding to the minimum labeled with letter B. These are the only modes that show promise for efficient energy transport through CROW bends (at least for bend angles around 20 and 50 degrees). As mentioned before, all the CROW supermodes appear as doublets in the CROW optical spectrum. Thus, every minimum in Fig. 5 corresponds to two closely located supermodes. Fig. 6 (a) and (b) show the change of the resonant frequencies and Q-factors of two anti-bonding $WG_{7,1}$ supermodes corresponding to dip B in Fig. 3. Plots in Fig. 6(a) reveal Q-factor enhancement of one of the modes in the doublet for the CROW bend angle of 22°. Similarly, the other mode is enhanced if the CROW bend angle is equal to 54°. Near-field portraits of these two modes are presented in Fig. 6(c) and demonstrate that the field distribution in the central resonator is far from that of the expected WG-mode field pattern. Instead, very low field intensity in this microdisk is observed, which can explain low bend radiation losses in such CROW configurations.

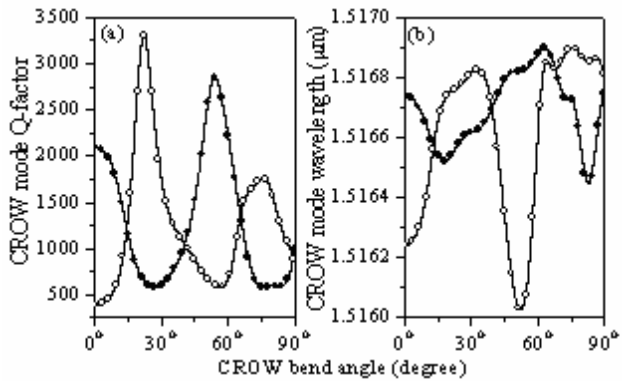

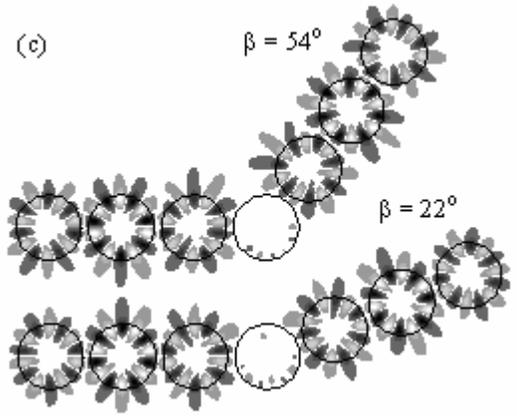

Fig. 6. (a) Quality factors change and (b) wavelengths migration of the B-doublet supermodes of the 7-resonator CROW with the same parameters as in Fig. 5a versus the value of the bend angle; (c) Near-field portraits of these supermodes for the CROW bend angles of 22° and 54°.

To check the validity of the prediction of the efficient transmission trough $2\pi/n$ bends in CROWs composed of larger weakly-coupled microdisks operating on the $WG_{n,m}$ modes, we consider a finite-size CROW made of 3.65 $\mu m$ radii resonators coupled via 500 nm airgaps. Transmission characteristics of such CROWs composed of 2,3...7 resonators are plotted in Fig. 7. Similarly to the previous case, we observe a number of dips in the normalized total scattered power curves, which correspond to the excitation of high-Q $WG_{20,1}$ supermodes in the CROW (dotted line indicates the value of the resonant wavelength of the $WG_{20,1}$-mode in an isolated resonator).

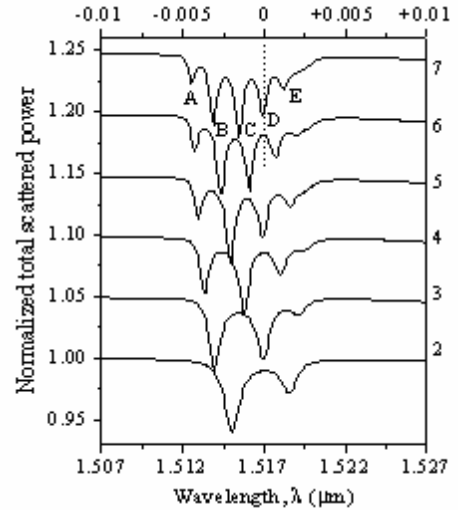

Fig. 7. Normalized total scattered power of a CPS beam scattering from a straight CROW section composed of identical microdisks with $a_i = 3.65\ \mu m$, $\varepsilon_i = 2.5$, $i = 1,...L-1$, $\varepsilon_L = (2.5, 0.001)$ as a function of the incident field wavelength for varying number of microdisks.

First, we consider the CROW supermodes corresponding to minimum C in Fig. 7 and study how bending the CROW changes their resonant wavelengths and Q-factors. The results presented in Fig. 8 (a) and (b) demonstrate the enhancement of the Q-factors of both modes in the doublet with the period of $2\pi/20^o = 18^o$. The supermodes near-field portraits are shown in Fig. 8(c) for bend angles of 27 and 54 degrees. Because the maxima of the plots in Fig. 7(a) are shifted by $9^o$, efficient transport through this CROW section can be achieved for the bend angles ranging from $0^o$ to $90^o$ with the increment of $9^o$. This provides design flexibility for making almost arbitrarily-bent CROWs.

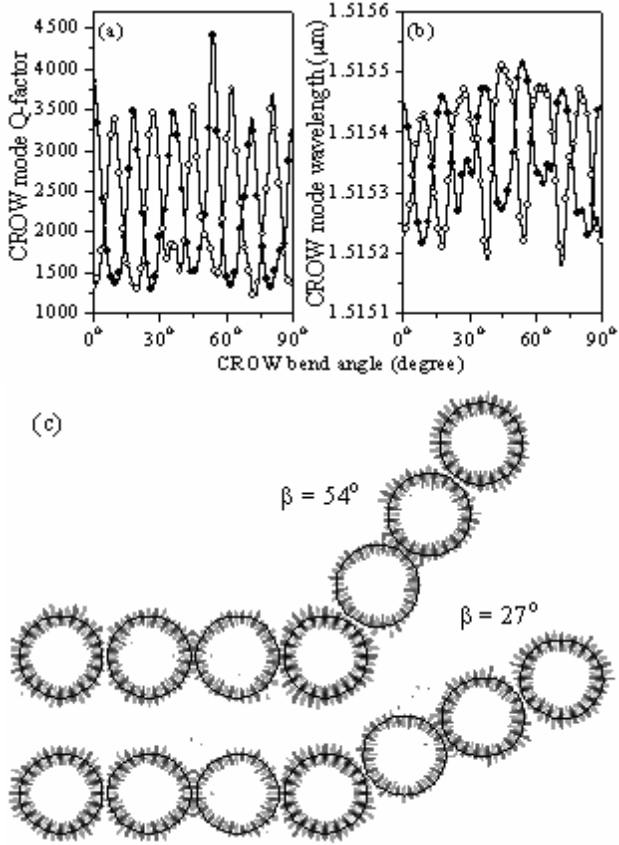

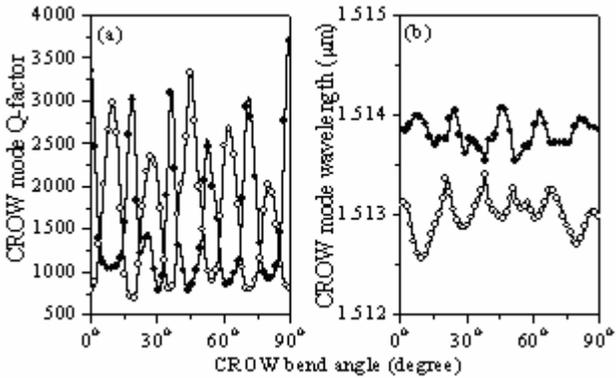

Fig. 8. (a) Quality factors change and (b) wavelengths migration of the C-doublet supermodes of the 7-resonator CROW with the same parameters as in Fig. 7 versus the value of the bend angle; (c) Near-field portraits of these supermodes for the CROW bend angles of $27^o$ and $54^o$.

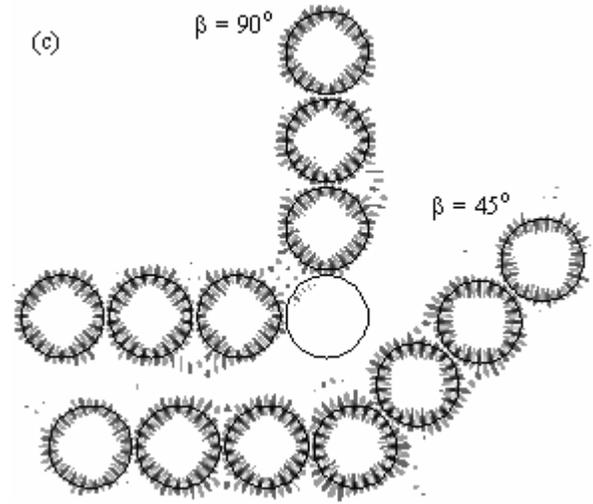

Fig. 9. Same as in Fig. 8 only for the the B-doublet CROW supermodes.

Next, to check the applicability of such design strategy to other high-Q CROW supermodes, we performed the same analysis for the modes corresponding to minimum B in Fig. 7. The results are presented in Fig. 9 and demonstrate a similar picture of the Q-factors enhancement of alternating modes with the period of $9^o$. Therefore, the previously formulated design rule of making loss-less CROW bends is indeed applicable to the case of weakly-coupled resonators operating on high-azimuthal-order WG modes. However, in sharply bent CROWs, the field pattern in the central resonator can be severely distorted from that of the whispering-gallery mode even in this case (see Fig. 9(c)).

## V. DESIGN OF LOW-LOSS BENDS IN CROWs COMPOSED OF WAVELENGTH-SCALE MICRODISK RESONATORS

It would be highly desirable to make possible efficient energy transfer along arbitrarily bent CROWs composed of strongly-coupled wavelength-scale microdisks. To achieve this goal, we propose to tune the size of the microdisk that is located at the CROW bend. Consider again the 7-resonator CROW studied in section IV. As it has been shown, efficient transmission in such a CROW can only be achieved through bends close to 22 and 54 degrees. We will now try to change the radius of the central resonator to achieve efficient transfer through arbitrarily-chosen bends, e.g., through 40- and 90-degree bends. As can be seen in Fig. 10, low bend losses can be achieved if the central disk has either smaller or larger radius than those of other disks. The field portraits of the corresponding CROW supermodes are presented in Fig. 10(e) for the values of the central disk radius that lead to noticeable enhancement of the supermodes Q-factors. These results demonstrate a possibility of making low-loss bends of arbitrary angles in CROWs composed of both weakly- and strongly-coupled wavelength-scale WG-mode microdisk resonators.

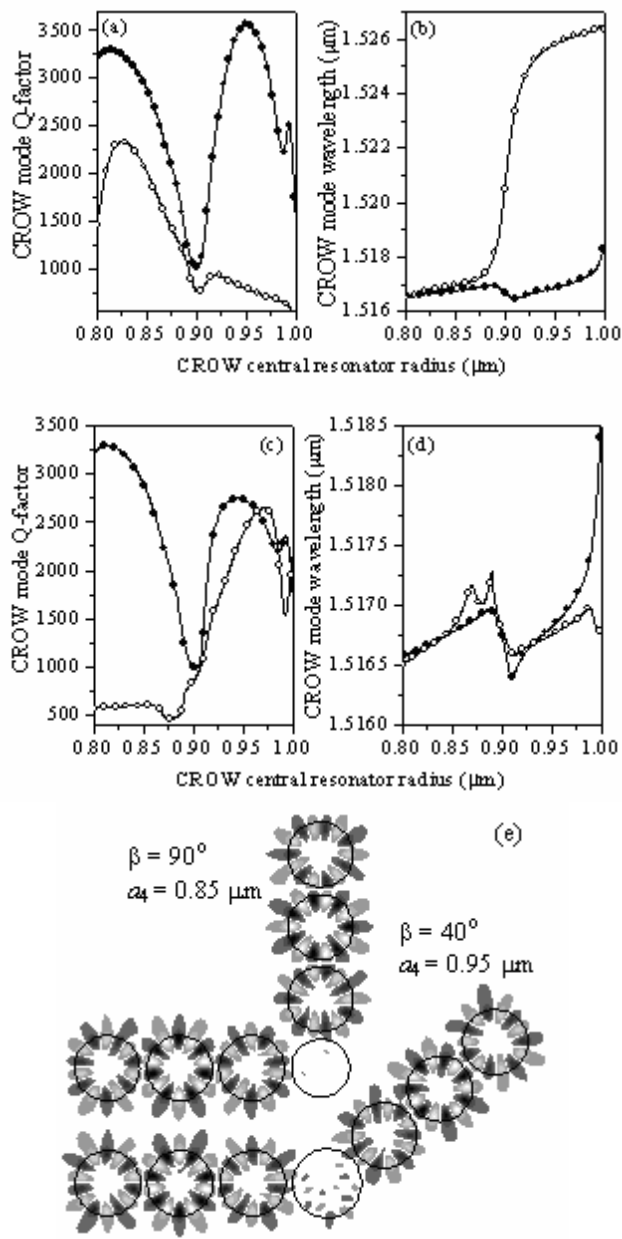

Fig. 10. (a,c) Quality factors change and (b,d) resonant wavelengths migration of B-doublet supermodes of the CROW with the same parameters as in Fig. 5a and Fig. 6 as a function of the central resonator radius for the bend angles of 40° and 90°, respectively. (e) Near-field portraits of the Q-enhanced supermodes.

## VI. CONCLUSIONS

Using the MBIEs formalism for the description of optical fields in finite-size aperiodic WG-mode microdisk CROWs, we demonstrate the power of this technique to reveal a detailed picture of CROW modal spectra and field profiles and thus to offer design strategies for controlled manipulation of the CROW transmission characteristics. It should also be noted that the method is applicable to modeling not only coupled-cavity structures with losses or gain but also to studying coupled-cavity plasmon waveguides composed of metal nanoparticles [32, 33].

Our results confirm the possibility of achieving efficient light transport in CROWs made of large weakly-coupled resonators through select bend angles that depend on the azimuthal order of the WG modes excited in the resonators (namely, through $\pi/n$-angle bends for WG$_{n,m}$ modes). For strongly-coupled wavelength-scale microdisks the picture is quite different; efficient light transport through arbitrary bends in not always possible in CROWs composed of identical resonators. However, we have proposed a general rule for achieving low-loss bends in such CROWs. Our design strategy is to either enlarge or reduce the radius of the microdisk located at the CROW bend. Our results also predict that in sharply-bent CROWs the WG-mode field distribution in the central resonator may become severely distorted, with regions of high field intensity appearing outside of the resonator material. This should be taken into account if other components are located on the optical chip in the close vicinity of the CROW bend region.

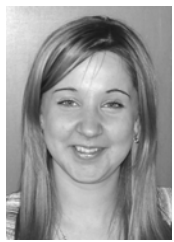

**Svetlana V. Pishko** was born in Kharkov, Ukraine in 1985. She received the B.Sc. and M.Sc. degrees in radiophysics from V. Karazin Kharkov National University, Kharkov, Ukraine in 2006 and 2007, respectively. Svetlana's current research interests focus on numerical simulation and design of electromagnetic and photonic components, including resonator and waveguide arrays.

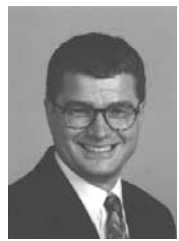

**Trevor M. Benson** (M'95–SM'01) was born in Sheffield, U.K., in 1958. He received the B.Sc. (with honors) degree in physics and the Ph.D. degree in electronic and electrical engineering in 1979 and 1982, respectively, both from the University of Sheffield, Sheffield, U.K. He received the D.Sc. degree in electrical and electronic engineering from the University of Nottingham, Nottingham, U.K., in 2005.

From 1983 to 1989, he was a Lecturer at the University College Cardiff, Cardiff, U.K. From 1989 to 1996, he was a Senior Lecturer in Electrical and Electronic Engineering, Reader in Photonics, and Professor of Optoelectronics. Currently, he is at the School of Electrical and Electronics Engineering, George Green Institute for Electromagnetics Research, University of Nottingham. His current research interests include experimental and numerical studies of electromagnetic fields and waves, with particular emphasis on propagation in optical waveguides and lasers, glass-based photonic circuits, and electromagnetic compatibility.

Dr. Benson is a Fellow of the Royal Academy of Engineering, the Institute of Electrical Engineers (IEE), and the Institute of Physics. He was the recipient of the Clark Prize in Experimental Physics from the University of Sheffield in 1979, and the *Electronics Letters* and J. J. Thomson Premiums from the IEE.

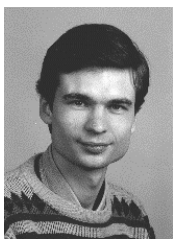

**Phillip D. Sewell** (S'88–M'91–SM'04) was born in London, U.K., in 1965. He received the B.Sc. (with honors) degree and the Ph.D. degree, both in electrical and electronic engineering, from the University of Bath, Bath, U.K., in 1988 and 1991, respectively.

From 1991 to 1993, he was a Science Education Resource Center Postdoctoral Fellow at the University of Ancona, Ancona, Italy. Since 1993, he has been with the School of Electrical and Electronic Engineering, George Green Institute for Electromagnetics Research, University of Nottingham, Nottingham, U.K., as a Lecturer, Reader, and Professor of Electromagnetics. His current research interests include analytical and numerical modeling of electromagnetic problems, with application to optoelectronics, microwaves, and electrical machines.

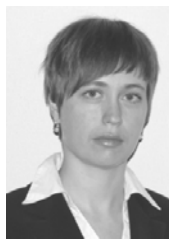

**Svetlana V. Boriskina** (S'96-A'99-M'01-SM'06) was born in Kharkov, Ukraine in 1973. She received the M.Sc. (with honors) degree in radiophysics and Ph.D. degree in physics and mathematics from V. Karazin Kharkov National University (KNU), Ukraine, in 1995 and 1999, respectively.

Since 1997 she has been a Researcher in the School of Radiophysics at KNU. From 2000 to 2005, she was a Royal Society – NATO Postdoctoral Fellow and a Research Fellow in the School of Electrical and Electronic Engineering, University of Nottingham, UK. Currently, she is a Senior Research Scientist in the School of Radiophysics at KNU. Her current research interests include computational electromagnetics, micro- & nano-photonics and plasmonics.

Dr. Boriskina authored and co-authored over 90 journal and conference publications. She has been awarded the SUMMA Foundation Graduate Fellowship in Advanced Electromagnetics, the IEEE MTT Society Graduate Student Scholarship, the Royal Society/NATO Postdoctoral Fellowship, the SPIE Educational Grant, and the ICO-ICTP Award.